\documentclass[11pt,twoside]{article}  % Leave intact
\usepackage{asp2006}
\usepackage{adassconf}

\newcommand{\pvmxstar}{{\tt pvm\_xstar}}
\newcommand{\pmodel}{{\tt PModel}}

\begin{document}   % Leave intact

\paperID{C.6}

\title{Parallelizing the XSTAR Photoionization Code}

%-----------------------------------------------------------------------
%          Short Title & Author list for page headers
%-----------------------------------------------------------------------
% Please supply the author list and the title (abbreviated if necessary) as 
% arguments to \markboth.
%
% The author last names for the page header must appear in one of 
% these formats:
%
% EXAMPLES:
%     LASTNAME
%     LASTNAME1 and LASTNAME2
%     LASTNAME1, LASTNAME2, and LASTNAME3
%     LASTNAME et al.
%
% Use the "et al." form in the case of four or more authors.
%
% If the title is too long to fit in the header, shorten it: 
%
% EXAMPLE: change
%    Rapid Development for Distributed Computing, with Implications for the Virtual Observatory
% to:
%    Rapid Development for Distributed Computing

\markboth{Noble et al.}{Parallelizing the XSTAR Photoionization Code}

%-----------------------------------------------------------------------
%		          Authors of Paper
%-----------------------------------------------------------------------
% Enter the authors followed by their affiliations.  The \author and
% \affil commands may appear multiple times as necessary.  List each
% author by giving the first name or initials first followed by the
% last name. Do not include street addresses and postal codes, but 
% do include the country name or abbreviation. 
%
% If the list of authors is lengthy and there are several institutional 
% affiliations, you can save space by using the \altaffilmark and \altaffiltext 
% commands in place of the \affil command.
%
% EXAMPLE: 
%      \author{Raymond Plante, Doug Roberts, 
%                  R.\ M.\ Crutcher\altaffilmark{1}}
%      \affil{National Center for Supercomputing Applications, 
%                 University of Illinois, Urbana, IL, USA}
%      \author{Tom Troland}
%      \affil{University of Kentucky, Lexington, KY, USA}
%
%      \altaffiltext{1}{Astronomy Department, UIUC}
%
% In this example, the first three authors, "Plante", "Roberts", and
% "Crutcher" are affiliated with "NCSA".  "Crutcher" has an alternate 
% affiliation with the "Astronomy Department".  The fourth author,
% "Troland", is affiliated with "University of Kentucky"

\author{Michael S. Noble, Li Ji}
\affil{Kavli Institute for Astrophysics,Massachusetts Institute of Technology}
\author{Andrew Young}
\affil{University of Bristol}
%\author{Li Ji}
%\affil{Kavli Institute for Astrophysics,Massachusetts Institute of Technology}
\author{Julia C. Lee}
\affil{Harvard University}

%-----------------------------------------------------------------------
%			 Contact Information
%-----------------------------------------------------------------------
% This information will not appear in the paper but will be used by
% the editors in case you need to be contacted concerning your
% submission.  Enter your name as the contact along with your email
% address.
% 
% EXAMPLE:  \contact{Dennis Crabtree}
%           \email{crabtree@cfht.hawaii.edu}

\contact{Mike Noble}
\email{mnoble@space.mit.edu}

%-----------------------------------------------------------------------
%		      Author Index Specification
%-----------------------------------------------------------------------
% Specify how each author name should appear in the author index.  The 
% \paindex{ } should be used to indicate the primary author, and the
% \aindex for all other co-authors.  You MUST use the following
% syntax: 
%
% SYNTAX:  \aindex{Lastname, F.~M.}
% 
% where F is the first initial and M is the second initial (if used). Please 
% ensure that there are no extraneous spaces anywhere within the command 
% argument. This guarantees that authors that appear in multiple papers
% will appear only once in the author index. Authors must be listed in the order
% of the \paindex and \aindex commmands.
%
% EXAMPLE: \paindex{Crabtree, D.}
%          \aindex{Manset, N.}        
%          \aindex{Veillet, C.}        

\paindex{Noble, M.S.}
%\aindex{}     % Remove this line if there is only one author

%-----------------------------------------------------------------------
%			Subject Index keywords
%-----------------------------------------------------------------------
% Enter up to 6 keywords that are relevant to the topic of your paper.  These 
% will NOT be printed as part of your paper; however, they will guide the creation 
% of the subject index for the proceedings.  Please use entries from the
% standard list where possible, which can be found in the index for the 
% ADASS XVI proceedings. Separate topics from sub-topics with an exclamation 
% point (!). 
%
% EXAMPLE:  \keywords{astronomy!radio, computing!grid, data management!workflows, 
%     instrumentation!control}


%-----------------------------------------------------------------------
%			       Abstract
%-----------------------------------------------------------------------
% Type abstract in the space below.  Consult the User Guide and Latex
% Information file for a list of supported macros (e.g. for typesetting 
% special symbols). Do not leave a blank line between \begin{abstract} 
% and the start of your text.

\begin{abstract}          % Leave intact
 We describe two means by which XSTAR, a code which computes physical
 conditions and emission spectra of photoionized gases, has been
 parallelized.  The first is \pvmxstar, a wrapper which can be
 used in place of the serial xstar2xspec script to foster concurrent
 execution of the XSTAR command line application on independent sets
 of parameters.  The second is \pmodel, a plugin for the Interactive
 Spectral Interpretation System (ISIS) which allows arbitrary components
 of a broad range of astrophysical models to be distributed across
 processors during fitting and confidence limits calculations, by
 scientists with little training in parallel programming.  Plugging
 the XSTAR family of analytic models into \pmodel\ enables multiple
 ionization states (e.g., of a complex absorber/emitter) to be computed
 simultaneously, alleviating the often prohibitive expense of the
 traditional serial approach.  Initial performance results indicate that
 these methods substantially enlarge the problem space to which XSTAR
 may be applied within practical timeframes.
\end{abstract}

%-----------------------------------------------------------------------
%			      Main Body
%-----------------------------------------------------------------------
% Place the text for the main body of the paper here.  You should use
% the \section command to label the various sections; use of
% \subsection is optional.  Significant words in section titles should
% be capitalized.  Sections and subsections will be numbered
% automatically. 
%
% EXAMPLE:  \section{Introduction}
%           ...
%           \subsection{Our View of the World}
%           ...
%           \section{A New Approach}
%
% It is recommended that you look at the sample paper sample2.tex
% for examples of formatting references, footnotes, figures, equations, 
% html links, lists, and other features.  

\section{Introduction}

XSTAR is ``a computer program for calculating the physical conditions and
emission spectra of photoionized gases" (Kallman \& Bautista 2001); the
science it facilitates may be described most concisely by paraphrasing
the documentation: {\em a
spherical gas shell surrounding a central source of ionizing radiation
absorbs some of this radiation and reradiates it in other portions of
the spectrum. XSTAR computes the effects on the gas of absorbing this
energy, and the spectrum of reradiated light, while allowing for
consideration of other sources (or sinks) of heat, such as mechanical
compression \&
expansion, or cosmic ray scattering.} Coded in Fortran 77, XSTAR may be
used as either a standalone executable or in the form of analytic models
like {\tt warmabs}, with the latter being compiled into shared objects
and dynamically loaded into spectral modeling tools such as ISIS
(Houck, 2002).
We are presently using XSTAR in ISIS to model active galactic nuclei
and non-equilibrium ionization of photoionized plasmas.  Relative
to classic
spectral modeling conducted with interactive analysis tools, the scales
of these efforts are large: analytic models with 20 or more components
\& roughly 300 parameters, scores of which may vary during fitting, or
batch XSTAR runs on thousands of individual sets of parameters.  The
compute time required in both use cases, on the order of a
week to a month for single end-to-end runs, precludes traditional use of
XSTAR, which is coded for serial execution on one CPU.  Compounding
the problem is the fact that most research efforts require multiple
end-to-end runs, e.g. to experiment with different model components or
parameter values, which can extend analysis timeframes into several
months.

\section{Batch Execution of XSTAR}
\begin{figure}[t]
\plotone{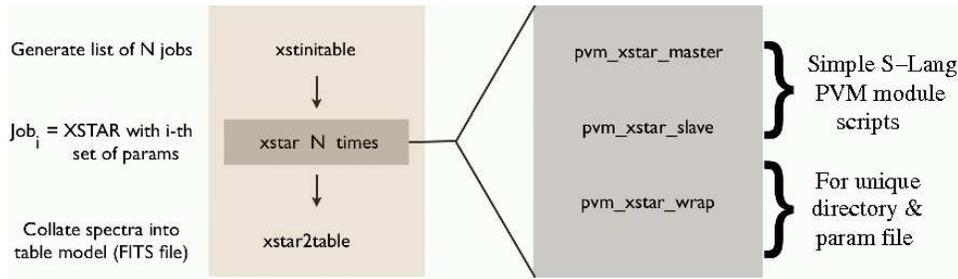}
     \caption{The flow diagrams of classic {\tt xstar2xspec} and its
     parallelized cousin, \pvmxstar, are identical (left): both run
     {\tt xstinitable} at outset and {\tt xstar2table} at completion.
     The only conceptual difference is that in \pvmxstar\ the N
     jobs are distributed to multiple CPUs via PVM (right), and executed
     in N unique directories to avoid FITS i/o \& parameter file clashes.
     }
\label{flow}
\end{figure}

Part of our non-equilibrium ionization modeling includes large-scale
simulations, wherein the XSTAR application is repeatedly invoked over
sets of unique input parameter tuples; one spectrum is generated per
XSTAR run and saved as a FITS file, and these are collated into a single
FITS table model that can be incorporated into an analytic model for
fitting.  Historically, this process has been driven by the serial
{\tt xstar2xspec} script bundled with XSTAR and outlined in
Fig. \ref{flow}.  A representative simulation of 600 XSTAR jobs,
generating power spectra of Hercules X1, consumed 26.4 hours of
wallclock time on a
% serial run was on garth-ranzz (see ~mnoble/conf/adass08/paper run logs)
single 2.6Ghz AMD Opteron processor with 2GB RAM; a linear scaling to
4200 jobs would consume 7.5 days on the same machine.  In contrast, a
similar physical simulation of 4200 XSTAR jobs completed in 110
minutes when executed via \pvmxstar\ on our Beowulf cluster of 52
2.4Ghz Opteron (4GB RAM) processors.  As shown in Fig. \ref{flow},
\pvmxstar\ consists of 4 scripts: 2 of these, \pvmxstar\ proper and
{\tt pvm\_xstar\_wrap}), are coded in Bourne shell, while the master/slave
scripts are coded in S-Lang using the S-Lang PVM module (Davis et al
2005, Noble et al 2006) to interface with the Parallel Virtual Machine
toolkit (Geist et al 1994).

\section{XSTAR Analytic Modeling}

As noted earlier, XSTAR is also used in the form of dynamically loaded
analytic models, as in this sequence of commands at the ISIS prompt:
{\small
\begin{verbatim}
        isis>  load_data("my_data.pha")
        isis>  model("warmabs(1) + warmabs(2) + hotabs(1)")
        isis>  set_params(...)
        isis>  fit

        Parameters [Variable]  = 48[21]
                    Data bins  = 3
                    Chi-square = 1.1118061
\end{verbatim}
}
The second step defines a 3-component model, consisting of two XSTAR
{\tt warmabs} components and one XSTAR {\tt hotabs} component.\footnote{Note
that in {\tt warmabs(1)} and {\tt warmabs(2)} the numbers
within parentheses are not parameters to the model, but rather are tags
which uniquely identify {\em instances} of a given model type, so that
each instance may be evaluated with its own set of parameter values.}
The performance bottleneck here is that each component may take 15 or more
seconds to evaluate %\footnote{Model component evaluation consists of invoking
%the underlying XSTAR Fortran function, with the parameters and grid defined
%in the third step of the command sequence given above.}
just once on a modern CPU, or
45 seconds to compute the entire model expression for every iteration
of the fit loop initiated by step 4.  A typical fit loop may contain
hundreds of such iterations, with tens of thousands to millions of
component evaluations often needed to conduct thorough walks through
parameter space while generating error bars.  In short, days or weeks of
compute time can be needed for essential analysis when expensive models
are involved.
\subsubsection{Latent Parallelism}
These lengthy runtimes may be shrunk by observing that there are two
sources of parallelism inherent to model evaluation. First, whenever
model components are mathematically independent of one another they
may be evaluated concurrently.  In the above model, for example,
each component may be evaluated simultaneously, potentially reducing
the runtime of each fit loop iteration from 45 to 15 seconds (the
theoretical maximum of linear speedup on 3 CPUs).  This component
independence is common in model expressions, which are evaluated from
left to right under the associativity and precedence rules
of classic algebra.  The second form of
parallelism arises from bin independence within models: when evaluating
the model on the {\em i}-th bin---\verb=model(lo[i], hi[i], params)=---requires
no knowledge of bins {\em i-1} or {\em i+1}, then wavelength/energy grids
of size {\tt nbins} may be trivially decomposed
{\small
\begin{verbatim}
      lo[1, nbins] = [ lo[1,N], lo[N+1, 2N] ... lo[nbins-N+1, nbins] ]
      hi[1, nbins] = [ hi[1,N], hi[N+1, 2N] ... hi[nbins-N+1, nbins] ]
\end{verbatim}
}
\noindent into {\tt nbins/N} subgrids and each
{\small\verb=model(lo_subgrid[j],hi_subgrid[j],params)=} evaluated
concurrently.  This is relatively common in models of X-ray spectra.

The \pmodel\ plugin for ISIS was written to exploit these latent sources
of parallelism.  Loaded at runtime by a simple \verb=require("pmodel")=
command, the package adds 4 primary functions to ISIS: {\tt pm\_add(),
pm\_mult(), pm\_func(), \& pm\_subgrid(N)}.
The first three are stub models, in that they contribute nothing to the
physics being modeled, but can be used in a model expression to identify
which portions to evaluate concurrently.  The fourth function is not a
stub model, but rather overrides the default model evaluation mechanisms
in ISIS with routines that decompose the model grid into N
independent subgrids.  In this case the entire model is independently
evaluated over pieces of the grid, while the first group of functions
evaluates pieces of the model independently over the entire grid.
Using \pmodel\ is easy: in the context of our XSTAR example only step 2
would need to change, to
{\small
\begin{verbatim}
        model("pm_add(warmabs(1), warmabs(2), hotabs(1))")
\end{verbatim}
}
For every iteration of the ISIS fit loop this revised model expression
would cause
the dispatch of each component evaluation to a distinct processor, with
the results from each combined by a simple additive {\em reduction}
operation.  Although \pmodel\ may be used to distribute virtually any
expensive model components, the same ease of use would apply: the parallel
use case bears an overwhelming resemblance to the serial one, with the
differences being simple to identify and implement.  %The simplicity of
%this approach
This means that end-users need
not learn to program for parallelism in order to use multiple processors
in their
models, a classic barrier to the adoption of parallel methods by
non-specialists.  The \pmodel\ functions will decompose the model or grid
and combine results with either additive, multiplicative, or
arbitrary functional reduction operations, all transparent to the
top-level user interface.  Moreover, ISIS did not
need to be recoded for parallelism, and in fact it does not even know
the model is computed in parallel; this knowledge is completely encoded
within \pmodel, whose functions ISIS simply calls in the same serial
manner it would for any other physical model component.
We have used these techniques to reduce the compute time of models with
20+ components, containing 10 or more XSTAR components and hundreds of
parameters, from 4+ weeks when run serially to \verb=~=22 hours on the
aforementioned Beowulf cluster.

\section{Conclusion}

Together, \pvmxstar\ and \pmodel\ enable scientists to incorporate multiple
processors in their XSTAR modeling without becoming experts in parallelism.
Amortizing the evaluation of expensive XSTAR components over many CPUs allows
larger and more physically realistic models to be computed, permitting us
to probe thousands
of physical scenarios in the time it has previously taken to compute only
a handful of such models.   Insofar as analytic modeling of observational
data is among the most common scientific activities in astronomy, \pmodel\
has a broad scope of applicability, particularly because it can in principle
distribute the evaluation of {\em any} expensive model, not merely the
XSTAR components shown here.  Both \pvmxstar\ and \pmodel\ are small open
source packages, and have been employed at several institutes, on multicore
desktops, workstation clusters, and high-performance parallel computers.
They may be obtained by download from
\url{http://space.mit.edu/cxc/pvm\_xstar/} or by contacting the lead author.

\acknowledgments
{\footnotesize
This work was supported by NASA through the Hydra AISRP grant NNG06GE58G,
and by contract SV-1-61010 from the Smithsonian Institution.}


\vspace*{-4mm}
\begin{references}
\footnotesize
\reference Kallman, T. \& Bautista, M.\ 2001, Photoionization and High-Density Gas, \apjs, 133, 221-253,
\reference Houck, J.~C.\ 2002, ISIS: The Interactive Spectral Interpretation System, High Resolution X-ray Spectroscopy with XMM-Newton and Chandra
\reference Davis, J.~E., Houck, J.~C., Allen, G.~E., and Stage, M.~D., 2005, \adassxiv, 444
\reference Noble, M.~S.,  Houck, J.~C., Davis, J.~E., Young, A., Nowak, M.\ 2006, Using the Parallel Virtual Machine for Everyday Analysis, \adassxv, 481
\reference A. Geist,  A. Beguelin,  J. Dongarra, W. Jiang, R. Manchek, \&
        V. Sunderam 1994, PVM: Parallel Virtual Machine, A User's Guide and Tutorial for Networked Parallel Computing

\end{references}
\end{document}